\documentclass
[12pt]
{article}
\usepackage{epsfig,latexsym,xspace,
url,subcaption}
\usepackage{a4,amsopn,
latexsym,
xspace,graphicx,hyperref,mathrsfs,epsfig,bm,bbm,enumitem,geometry,marginnote}
\newcommand{\CIP}{\mathop{\perp\!\!\!\perp}}
\newcommand{\pa}{{\rm pa}}

\usepackage[numbers]{natbib}
\newcommand{\textcite}{\citet}

\title{Potential Outcomes and Decision Theoretic Foundations for Statistical Causality: Response to Richardson and Robins}
\author{A.~Philip~Dawid\thanks{University of Cambridge}}

\newcommand{\eg}{{\em e.g.\/}\xspace}

\newcommand{\etc}{{\em etc.\/}\xspace}

\newcommand{\HIDE}[1]{ }
\newcommand{\COMMENT}[1]{ }

\newcommand{\eqref}[1]{\mbox{(\ref{eq:#1})}}

\newcommand{\secref}[1]{\mbox{\S$\,$\ref{sec:#1}}}

\newcommand{\lemref}[1]{\mbox{Lemma~\ref{lem:#1}}}

\newcommand{\defref}[1]{\mbox{Definition~\ref{def:#1}}}
\newcommand{\fnref}[1]{\mbox{footnote~\ref{fn:#1}}}

\newcommand{\corref}[1]{\mbox{Corollary~\ref{cor:#1}}}

\newcommand{\remref}[1]{\mbox{Remark~\ref{rem:#1}}}

\newcommand{\Secref}[1]{\mbox{Section~\ref{sec:#1}}}

\newcommand{\parents}[1]{{\rm pa}(#1)}

\newcommand{\pre}{{\rm pre}}

\newcommand{\implies}{\;\Rightarrow\;}

\newcommand{\cip}{\mbox{$\perp\!\!\!\perp$}}
\newcommand{\indo}[2]{\mbox{$#1 \,\cip\, #2$}}
\newcommand{\ind}[3]{\mbox{$#1 \, \cip\, #2 \mid #3$}}

\newtheorem{expl}{Example}
\newtheorem{definer}{Definition}

\newtheorem{algor}{Algorithm}
\newtheorem{lemma}{Lemma}
\newtheorem{lem}[lemma]{Lemma}

\newtheorem{cor}{Corollary}
\newtheorem{rem*}{Remark}

\newcommand{\halm}{\hspace*{\fill} $\Box$\par}
\newenvironment{proof}{\noindent {\bf Proof. }}{\halm\vspace{\baselineskip}}

\newcommand{\idle}{\mbox{$\emptyset$}}

\begin{document}\maketitle

\begin{abstract}
  \noindent
  I thank Thomas Richardson and James Robins for their discussion of
  my paper, and discuss the similarities and differences between their
  approach to causal modelling, based on single world intervention
  graphs, and my own decision-theoretic approach.
  \\[1ex]
\end{abstract}

\noindent 

\section{Introduction}
\label{sec:intro}

I am indebted to 
\textcite{RandR:discussion}, henceforth RR, for their serious and
detailed engagement with the ideas and material in my paper
(\textcite{apd:found}, henceforth D21). It is particularly valuable
that they highlight the similarities and differences between my
decision-theoretic (DT) approach and their own approach based on
Single World Intervention Graphs (SWIGs). Indeed the similarities are
manifold, and the differences few and largely inconsequential. I will
however concentrate here on these small differences, in the hope that
this will illuminate the differences in our underlying world views.

In \secref{spec} I address some specific points raised by RR's
discussion, and in \secref{crit} I respond to various critiques they
make of D21.  \Secref{alt} addresses RR's argument, an alternative to
the one I gave in D21, and supplies some corrections to their
analysis.  Finally in \secref{disc} I opine on the relative advantages
and disadvantages of SWIG and DT representations.

\paragraph{Note:}
In the sequel, references to equations in RR are given in the form
``equation~(R1)'', to those in D21 in the form ``equation~(D1)'', and
to those in the present article in the form ``equation~(1)'', with a
similar convention for other references.

\section{Some specific points}
\label{sec:spec}
\begin{enumerate}

\item RR's introduction says that I
  \begin{quote}
    {\em aim to develop a graphical framework} for causal models.
  \end{quote}
  Not exactly.  The fundamental idea of DT is that we can express
  causal properties by means of extended conditional independence
  (ECI) assertions, involving both stochastic variables and
  non-stochastic intervention indicators; and my paper aimed to
  develop arguments to support such assertions.  These arguments can
  always be expressed and developed non-graphically, using the purely
  algebraic theory of ECI.  It is true that graphical representations
  are incredibly useful and near-ubiquitous, which is why I devoted
  much attention to them in my paper; but the underlying theory does
  not require that we have such a representation (which is in any case
  not always available).
\item Footnote~R1.  I am disappointed that RR choose to perpetuate the
  prevalent but highly misleading terminological confusion between
  concepts that relate to distinct rungs of the ``ladder of
  causation'' \cite{pearl:why}.  Use of the same term
  ``counterfactual'' to denote totally distinct things is particularly
  dangerous, and I have often found myself confused, in reading the
  literature, as to which concept is intended.  A recent workshop I
  attended was entitled ``Counterfactual Prediction'', but was simply
  about using data to make forecasts for new patients, under various
  treatments.  Nothing about this runs counter to known facts, and it
  was thus firmly positioned on Rung~2 of the causal ladder, which
  concerns the effect of a new, actual or hypothetical, intervention
  on a system---so not involving a contradiction with any known facts,
  and not meriting the description ``counterfactual''.  On rung~3, by
  contrast, we ask genuinely counterfactual questions about what might
  have happened in a particular case if---in contradiction to the
  known facts---an action other than the actual one had been
  performed.  The contrast between these tasks is illuminated in
  \textcite{apd:kent,apd/mm:eoccoe}, where it is shown that different
  mathematical frameworks are required to formalise them.  In
  particular, while potential outcomes can be used at either of these
  levels, they are totally inessential for rung~2---for all that this
  accounts for by far the largest share of their current use---but
  seem unavoidable for rung~3.

  Both my own paper and that of RR stand firmly on rung~2, and involve
  no genuinely counterfactual considerations.  That is why I have been
  able to dispense entirely with potential outcomes, while still
  having a theory that---as RR convincingly show---is essentially
  isomorphic to theirs, where they have opted to employ them.

\item In \S{R2}, referring D21's ``hypothetical distributions'', RR
  say:
  \begin{quote}
    {\em there is no requirement that these distributions live on the
      same probability space.}
  \end{quote}
  The various regime distributions all relate to identical variables,
  and thus do live on a single space, though admittedly it is not not
  under the control of a single probability measure, so not a
  probability space.  This is the same structure we are familiar with
  in the context of a parametric statistical model.
  
  In RR's approach, each variable is indexed with one or more actions,
  leading to a proliferation of variables.  These variables can, if so
  desired, be considered as having an overall joint distribution---so
  ``living on the same probability space''; but it is only margins of
  this joint distribution, which are just my regime distributions,
  that are relevant.  In particular, the dependence, in the overall
  joint distribution, between versions of the same variable labelled
  by different interventions is both unknowable and (fortunately!)
  irrelevant.  So the advantage of having a single probability space
  is lost on me.  A similar approach, if applied for a parametric
  statistical model for a variable $X$ with parameter $\theta$, would
  involve constructing an expanded collection of variables
  $\{X(\theta)\}$, one for each value of $\theta$, all having a joint
  distribution---of which only the margins are of interest.  Why would
  one ever do such a thing?
  
\item Also in \S{R2} they say:
  \begin{quote}
    {\em Owing to the fundamental problem of causal inference the
      resulting factual distribution is consistent with many different
      intervention distributions.}
  \end{quote}
  This ambiguity is not related to the so-called ``fundamental problem
  of causal inference''\footnote{which in reality is not fundamental
    at all, but a massive own goal for the potential outcome
    formulation of causal inference.  It simply does not arise in a DT
    approach.} \cite{pwh:jasa}.  Rather, as a matter of logic, there
  is no necessary relation between how a system behaves when it is
  being observed, and how it behaves when it is kicked.

\item Section~{R3}: Labelling issues and individual effects.

  RR describe three distinct ways---uniform, temporal, and causal---in
  which variables in SWIGs may be labelled by actions.  They say
  \begin{quote}
    \em although we may wish to adopt the additional equalities
    between potential outcomes that are implied by the temporal and/or
    causal relationships, our results do not require these equalities
  \end{quote}
  ---indicating that it really does not make any difference which
  scheme is employed,

  RR opt to work with uniform labelling.  Indeed, either of the other
  schemes would not be representing a single world.  Thus, as they
  point out, Figure~R1(d) represents the case that $C$ would take the
  same value in the distinct worlds corresponding to actions
  $(a=0, b=1)$ and $(a=1, b=1)$.  While such ``absence of individual
  effects'' assumptions may have some intuitive appeal, they add
  nothing to the analysis.

  My personal view is that the very the concept of an ``individual
  effect'' is not merely unnecessary but metaphysical---and not in a
  good way \cite{apd:cinfer}. In particular, the emphasis, in the
  potential outcome approach, on necessarily
  unknowable\footnote{because of the ``fundamental problem of causal
    inference''} individual effects, takes one down a
  blind alley, which one has to re-emerge from before anything useful
  can be done.
  
\item Footnote~R12:
  \begin{quote}
    {\em In Dawid (2021, Figure 15), two conditions are stated as
      supporting $g$-computation.  The first of these is correct, but
      the second should be $\indo {Y{(x_0, x_1)}} {X_0}$, not
      $\indo {Z(x_0)} {X_0}$.}
  \end{quote}  

  I gratefully accept RR's correction, based on
  \textcite{jr:cma}.\footnote{See \textcite{dd:ss}, \S{8}, for a
    general decision-theoretic formulation and analysis, and \S10.2
    for its relation to the potential response approach of
    \textcite{jr:cma}.}  This requires the following amendments to
  D21: \setcounter{equation}{77}
  \begin{description}
  \item[Equation~(D78)]  Replace by:
    \begin{displaymath}
      \label{eq:17}
      \indo {Y(x_0,x_1)} {X_0}.
    \end{displaymath}
\setcounter{equation}{80}
    
\item[Equation~(D81)] Replace by:
  \begin{displaymath}
    \label{eq:18}
      \ind Y {X_0^*} {(F_0=x_0,F_1=x_1)}
    \end{displaymath}
    (noting that the dotted arrow from $X_0^*$ to $X_0$ disappears).

  \end{description}
  \setcounter{equation}{0}
   
\item Footnote~R20: I thank RR for catching my careless error.
   
\item Section~R6: The r\^ole of fictitious independence.

  Note that instead of equation~(R57) I had $\CIP_{i=1}^k F_i$, which
  is a preferred notation.

  I stand corrected by RR's analysis here, and am indeed embarrassed
  that I myself have fallen foul of the very fallacy I identified and
  analysed in \textcite{apd:misl} and \textcite[\S8.1]{apd:ciso}.
  Fortunately this is easily rectified with an additional assumption
  such as in \S{R6.2}, or, more straighforwardly (if more
  restrictively) {\em variation independence\/}, as described in
  \remref{1} below.  In partial deflection of their criticism I point
  out that (as mentioned in \remref{1}; see also \fnref{1}) RR also
  rely on just such an implicit assumption.
\end{enumerate}

\section{Response to RR's `Critique of Dawid's Proposal'}
\label{sec:crit}

\subsection{\S{R4.1}}
\label{sec:4.1}
RR argue against my aim of expressing causal properties by means of
augmented DAGs (or, more generally, extended conditional independence
statements) including regime indicators but without ITT variables.
They ask ``why it is necessary to introduce the ITT variables in the
first place?''

The first point they make in this Section was previewed in their
Introduction:
\begin{quote}
  {\em ITT variables are necessary and important in order to encode
    the notion of ignorability and the effect of treatment on the
    treated.}
\end{quote}
But this is not so.

\begin{description}
\item[Ignorability] In the DT approach, ignorability
  is directly encoded by extended conditional independence, \eg\
  $\ind Y {F_T} T$, without any need to consider ITT variables.  The
  sole purpose of introducing ITTs in D21 was to supply one possible
  argument that (when appropriate) might be made in justification of
  such assertions---and thus to justify (when appropriate) the use of
  an augmented DAG, without explicit ITT nodes, to represent and
  manipulate causal relations.

\item[Effect of treatment on the treated (ETT)] While consideration of
  ITTs is one way of thinking about ETT, it is not essential.  ETT can
  be meaningfully and helpfully interpreted in ways that do not
  involve ITTs at all \cite[\S34.4 and \S34.5.1]{sgg/apd:ett}.
\end{description}

RR go on to argue, by means of an example, that an augmented DAG,
without ITTs, can not distinguish a ``genuine'' causal relationship
from a ``spurious'' one.  But, as with any model, it is essential to
keep in mind the real-world characteristics that the ingredients of
the model are intended to represent.  In particular, the states of the
decision node index the data distributions associated with carefully
described hypothesised interventions.


The ``spurious'' case they discuss is represented\footnote{Well, not
  entirely, since the assumed underlying distribution is unfaithful to
  the graph, having the property $\ind Y {F_T} T$ even though that is
  not represented in it.}  by Figure~R4(b).  This involves particular
``fat hand'' interventions, and describes their effect on the response
$Y$.  In the---admittedly implausible---case that I myself was
considering undergoing just such fat-hand interventions, this might
describe my own decision problem.  And it would then indeed be the
case (assuming I could accept the appropriate exchangeability
assumptions) that I could consider the observational distributions of
$Y$ given $T$ as germane to that problem, and so alternatively
represent the problem as in Figure~R4(a).  As RR say, correctly, ``the
causal diagram shown in (a) cannot be refuted''.  But in this case, in
Figure~R4(a) the states of $F_T$ would represent the ``fat hand''
interventions, and, with this interpretation, the problem can still
appropriately be called ``causal''.  Such a representation would be
distinguishable from what RR consider to be a ``genuine'' causal case,
likewise represented by Figure~R4(a), but where the states now
represent different, ``surgical'', interventions.

As mentioned earlier, the introduction of ITT variables in D21 was
made to support ignorability assertions, here $\ind Y {F_T} T$, in
particular kinds of problems.  But they are not essential, and all
that RR's example demonstrates is that ignorability can hold even when
the argument based on ITTs fails.  This does not make such a problem
any less genuinely causal.  Moreover, while the decision-theoretic
description is unproblematic, I do not see how a SWIG approach could
represent ignorability in such a problem.

\subsection{\S{R4.2.1}}
\label{sec:4.2.1}
Here RR argue that I could (should?) have regarded variables such as
$T^*$ appearing in different regimes as identical, not merely
identically distributed.  But as I mentioned above when discussing
labelling issues, {\em while ``absence of individual effects''
  assumptions may have some intuitive appeal, they add nothing to the
  analysis\/}.

I am bemused by RR's complaint that my DT account ``leads to an
unnecessary multiplicity of random variables'', when in their SWIG
approach, even using the relatively lean causal labelling scheme,
every single variable is replaced by a host of potential variables
(one for each combination of actions that could affect it).

\subsection{\S{R4.3}}
\label{sec:4.3}

While, as I argued in D21, it is extremely useful to think about ITT
variables when trying to justify ignorability assumptions, I do not
agree with RR's preference to retain the intention-to-treat variable
$T^*$ in the final augmented DAG representation, while omitting the
received treatment variable $T$.  First, as I have already argued,
$T^*$ is not needed to ``rule out spurious invariance'', since this is
not in fact spurious; nor is $T^*$ essential for defining the effect
of treatment on the treated. (I do not rule out that there may be some
special cases where it is helpful to retain $T^*$, as well as $T$, in
the final model---in which case by all means include it also.)

Secondly, again as mentioned in the case of ``spurious causation'', in
some cases properties such as ignorability can be meaningfully
justified, and again expressed by $\ind Y {F_T} T$, even when no
argument involving ITT variables is available---in such a case $T$ is
essential, while $T^*$ is a red herring.

Another case is where causality is understood as a property of
invariance across differing contexts \cite{bhlmann2018invariance}, as
in considerations of transportability and external validity
\cite{pearl/bareinboim:sts}.  For example, a medical device may have
the same probability of registering a positive result, given whether
or not a patient has a certain condition, irrespective of who it is
used on, or in which hospital.  This can still be encoded as
$\ind Y {F_T} T$, where $Y$ denotes the response, $T$ the condition,
and $F_T$ the context.  Considerations of ignorability and
intention-to-treat are simply not relevant here, and there is no SWIG
representation.

\subsection{\S{R5}}
\label{sec:5}
At (R36), and again in Figure~R7, RR point out that a contextual
independence, here $\ind Y {F_T} {M, F_T\neq \idle}$, is not implied
by the (non-contextualised) conditional independence assumptions that
they label A and B.  This is so, but I can't see why it is a problem.
The full set of required assumptions includes, as well as A and B, the
description, in (D33), of how $T$ depends on $T^*$ and $F_T$.  Taken
all together, these imply all relevant contextual independencies.
Moreover the inclusion of dashed edges in an augmented DAG allows such
properties to be derived directly from the graph, bypassing algebraic
manipulations.

\section{RR's alternative argument}
\label{sec:alt}
The major part of RR is devoted to an argument alternative to the one
that I presented in Section~2 and Appendix~A of D21---using different
assumptions and arguments, but leading to the same conclusion.  RR
very helpfully conduct this argument twice, first (in
Sections~R3.2--R3.7) in the language of SWIGs, and again (in
Sections~R5.1--R5.6) in my own decision-theoretic language.  I like
this alternative development, and especially appreciate the two
parallel descriptions: it is illuminating to compare different ways of
looking at the same thing.  In particular the twin analyses
demonstrate the close correspondence between our approaches, such
differences as there are being largely (though not entirely)
notational.

The argument presented by RR appears basically correct, but certain
details of it, particularly in its decision-theoretic version, require
clarification and amplification.

\subsection{Distributional consistency}
\label{sec:DC}
Following their introduction of their own version of ``distributional
consistency'' in Definition~R2, RR give a variation on this definition
in terms of a ``dynamic regime'' $g_i^*$, supposed to have the effect
of setting an intervention target, $B_i$ (in my unstarred notation),
to agree with its ``natural value'' (which I conceive of as an
``intention-to-treat'', ITT, variable, $B_i^*$).  I do not see how
this advances the argument.  In particular, it generates still further
proliferation of potential variables, which now require $g_i^*$ as an
additional argument.  And in order for this to work at all, a variety
of additional conditions, as detailed in footnote~R8, are required,
detracting considerably from this approach---which is in any case
totally superfluous.

When RR introduce distributional consistency in the decision-theoretic
context, in Definition~R13, they do so solely in terms of $g_i^*$.
But the description of what $g_i^*$ does is indistinguishable from how
the idle regime $\idle$ operates.  So the two states of $F^*$ embody a
distinction without a difference and collapse into one---making the
interpretation of (R38) problematic,\footnote{In any case, (R38) is
  incomplete as formulated, since it does not specify the values of
  the regime indicators $F_{A\setminus (C\cup B_i)}$.  Presumably,
  here and elsewhere in RR, any unmentioned regime indicators are
  implicitly supposed idle.}  and rendering the argument from (R39) to
(R40) decidedly dodgy.  In particular, the first equality again
requires additional conditions, translations of those in footnote~R8,
which effectively beg the question.  Fortunately variables such as
$g_i^*$ are, again, entirely superfluous and---as RR themselves later
acknowledge---decision-theoretic distributional consistency can
perfectly well be defined by the equality of (R39) and (R40), which I
phrase as:\footnote{Henceforth, unless otherwise indicated, I follow
  RR in dropping the ${}^*$ notation, and consider only the ITT
  variables associated with intervention targets.}
\begin{definer}[Distributional consistency]
  \label{def:dc}
  This requires that, for $B\in A$, $Y=V\setminus B$,
\begin{equation}
  \label{eq:1}
  \Pr(Y=y, B=b \mid {F_B=b, F_{A\setminus B}}) = \Pr(Y=y, B=b \mid
  F_B=\idle, F_{A\setminus B})
\end{equation}
(where we do not distinguish between a variable and its singleton
set).
\end{definer}
\defref{dc} is the direct DT translation of the SWIG-based
Definition~R2.  Note that, in \eqref{1}, we could set the value of
$F_{A\setminus B}$ as
$(F_C=c\neq\idle, F_{(A\setminus B)\setminus C}= \emptyset$), for
$C\subseteq A\setminus B$, showing more clearly the equivalence with
(R39) and (R40).  This simplification will be used without further
comment in the sequel.

\begin{rem*}
  \label{rem:1} Note that it is implicitly assumed, here and in the
  sequel, that knowing the values (fixed or idle) of some intervention
  indicators (here $F_{A\setminus B}$) does not constrain the possible
  values of others (here $F_B$)---the property of {\/\em variation
    independence} \cite{apd:varind}.  This assumption---or a suitable
  weaker one, such as in (R64)---is also required throughout RR's
  arguments in \S{R3} for SWIGs,\footnote{where variation independence
    is equivalent to the existence of $p(V(a))$ in Equation~(R1), for
    all $a\in{\cal X}_D$, for each subset $D$ of $A$.} as well as
  their decision-theoretic \S{R5}.
\end{rem*}

\subsubsection{Relationship with D21's distributional consistency}
\label{sec:restatement}

Equation~\eqref{1} is equivalent to the pair of properties:
\begin{eqnarray}
  \label{eq:dc2}
  \Pr(Y=y \mid B = b, F_B=b, F_{A\setminus B}) &=&     \Pr(Y=y \mid B = b, F_B=\idle, F_{A\setminus B})\\
  \label{eq:dc1}
  \Pr(B = b \mid F_B=b, F_{A\setminus B}) &=&  \Pr( B = b \mid F_B=\idle, F_{A\setminus B}).
\end{eqnarray}
    
Equation~\eqref{dc2} is similar to my own definition of distributional
consistency (Definition~D2), applied under given interventions on some
or all of the variables in ${A\setminus B}$.  A difference is that I
was implicitly considering $Y$ to comprise ``response variables'' that
could be affected by $B$, whereas RR also allow variables that are
causally prior to $B$.  This seems very reasonable, and indeed
essential if we have not yet introduced a causal ordering of the
variables.
  
As for \eqref{dc1} (not in itself a ``distributional consistency''
property): because it involves the same value $b$ for both $B$ and
$F_B$ on the left, it is a weaker\footnote{unless $B$ is binary}
version of
\begin{equation}
  \label{eq:2}
  \ind {B} {F_B} {F_{A\setminus B}},
\end{equation}
which extends \eqref{dc1} to allow $F_B = b' \neq b$ on the left.
Condition~\eqref{2} encodes the intuitively desirable property that
(for any interventions on some or all of the other manipulable
variables) the distribution of the ITT variable $B$ (which is $B^*$ in
my own notation) is not affected by applying any intervention, or
none, to its target (my unstarred $B$).  Extending a remark in
footnote~R6, under the conditions of Lemma~R8 the stronger property
\eqref{2} will in any case hold.

\subsubsection{Lemma~R14}
\label{sec:rr14}
Because of its reliance on $F_B^*$, Lemma~R14 is meaningless as
stated.  Its statement and proof should be replaced by a DT paraphrase
of Lemma~R3, as follows:
\begin{lemma}
  \label{lem:rr14}
  Distributional consistency implies that \eqref{1} continues to hold
  for $B$ a general subset of $A$ and $Y \subseteq V\setminus B$.
\end{lemma}

\begin{proof}
  Use induction on the cardinality of $B$.  Write $B$ as a disjoint
  union $B = D \cup E$, with $E$ a singleton.  Then, with $Z = V\setminus B$,
\begin{eqnarray}
 \nonumber \lefteqn{\Pr(Z=z, B=b \mid F_B=b, F_{A\setminus B})}\\
 \nonumber &=& \Pr(Z=z, D=d, E= e \mid F_{D}=d, F_{E}= e,  F_{A\setminus B})\\ 
\label{eq:l3}  &=& \Pr(Z=z, D=d, E= e \mid F_{D}=d, F_{E}= \idle, F_{A\setminus B})\\
  \label{eq:l4} &=& \Pr(Z=z, D=d, E= e \mid F_{D}=\idle,  F_E=\idle, F_{A\setminus B})\\
 \nonumber &=& \Pr(Z=z, B=b \mid F_B=\idle, F_{A\setminus B}).
\end{eqnarray}
Here \eqref{l3} follows from \eqref{1}, and \eqref{l4} by the
inductive hypothesis.  Finally marginalize from $Z$ to $Y$.
\end{proof}

\subsection{The further argument}
\label{sec:further}
While the results in the remainder of \S{R5} are essentially correct,
there are some deficiencies in the arguments employed.

\subsubsection{Lemma~R15}
\label{sec:rr15}
Again because of it reliance on $F_B^*$, Lemma~R15 is meaningless as
stated.\footnote{However this lemma does not appear to be used by RR
  in the sequel.}  A suitable DT translation of Lemma~R4 is
\begin{lemma}
  \label{lem:rr15}
  Let $B \subseteq A$, and $W\subseteq Y=V\setminus B$.  Then under
  distributional consistency,
    \begin{equation}
      \label{eq:3}
      \Pr(Y = y \mid B = b, W, F_B=b, F_{A\setminus B}) = \Pr(Y = y \mid B = b, W, F_B=\idle, F_{A\setminus B}).
    \end{equation}
\end{lemma}
This follows directly on further conditioning \eqref{dc2} on $W$.

\subsubsection{Lemma~R16}
\label{sec:rr16}
The introduction of $F_B^*$ in the proof of Lemma~R16 is pointless:
the passage from line~2 to line~5 is immediate from distributional
consistency expressed as the equality of (R39) and (R40).  To clarify,
I re-express Lemma~R16 and its proof as follows, where by annotating
an intervention variable with the check mark $\check{}$ we understand
that it does not take value $\idle$.\footnote{In D21 it was the
  variable itself, rather than its intervention variable, that was so
  annotated.}

\begin{lemma}
  \label{lem:rr16}
  Let $B\subseteq A$ and $W\supseteq B$.  Then under
  distributional consistency
  \begin{equation}
    \label{eq:7}
    \ind W {\check F_B} {F_{A\setminus B}}\implies \ind W {F_B} {F_{A\setminus B}}.
  \end{equation}
\end{lemma}

\begin{proof} Let $w$ be a possible state of $W$, with projections
  $w', w''$ onto $W\setminus B, B$ respectively.  Then
  \begin{eqnarray}
    \label{eq:65}
    \lefteqn{\Pr(W=w \mid \check F_B = b, F_{A\setminus B})}\\
    \label{eq:68}   &=& \Pr(W\setminus B =w', B = w'' \mid \check F_B = b,
                    F_{A\setminus B})\\
    \label{eq:6}
                &=&  \Pr(W\setminus B =w', B = w'' \mid \check F_B = w'',
                    F_{A\setminus B})\\
    \label{eq:66} &=&  \Pr(W\setminus B =w', B = w'' \mid F_B = \idle,
                       F_{A\setminus B})\\
    \label{eq:67}
                &=&    \Pr(W = w \mid F_B = \idle,
                    F_{A\setminus B}).
  \end{eqnarray}
  Here \eqref{6} follows from the premise of \eqref{7}, and \eqref{66}
  from \lemref{rr14}.
\end{proof}

The next result is not explicit in RR, but is useful.
\begin{cor}
  \label{cor:aa}
  Further let $D\subseteq A$ be disjoint from $B$. Then
  \begin{equation}
    \label{eq:9}
    \ind W {(\check F_B, F_D)} {F_{A\setminus (B\cup D)}}\implies \ind
    W {(F_B, F_D)} {F_{A\setminus (B\cup D)}}.
  \end{equation}
\end{cor}

\begin{proof}
  Fix $w$.   The premise of \eqref{9} implies that there exists a variable $G$,
  measurable with respect to $A\setminus (B\cup D)$, such that, for any
  value $b\neq\idle$ of $F_B$,
  \begin{equation}
    \label{eq:10}
    \Pr(W=w \mid F_B=b, F_D, F_{A\setminus (B\cup D)}) = G.
  \end{equation}
  Then the equality of \eqref{65} and \eqref{67} shows that \eqref{10}
  holds also for $b=\idle$, and the result follows.
\end{proof}

\begin{cor}
  \label{cor:bb}
  \lemref{rr16} and \corref{aa} continue to hold if some or all of the
  intervention indicators $F_i$ in $F_{A\setminus B}$ are replaced by
  their checked versions $\check F_i$.
\end{cor}

\subsubsection{Lemma~R17}
Again, the line in the proof of Lemma~R17 involving $F_B^*$ is
superfluous and should be omitted.  Our version of its statement is as
follows.

\begin{lem}
  \label{lem:rr17}
  Let $B\subseteq A$, and let $Y$ and $W$ be disjoint with $B\subseteq
  W$.   Then under distributional consistency,
  \begin{equation}
    \label{eq:8}
    \ind Y {\check F_B} {(W, F_{A\setminus B})} \Rightarrow \ind Y {F_B} {(W, F_{A\setminus B})}.
  \end{equation}
\end{lem}
The proof is similar to that of \lemref{rr16}.

\subsubsection{Lemma~R19}
RR's proof of Lemma~R19 is inadequate in a number of ways.

\begin{description}
\item[(R48)] RR's argument for equating (R47) and (R48) fails because
  the property
  \begin{displaymath}
    \ind {W_{\pre(i)\cup\{i\}}} {F_{A\setminus \pre(i)}}
      {F_{A\cap\pre(i)}, F_A \neq\idle}
    \end{displaymath}
    does not satisfy the requirement ``$B\subseteq W$'' that would
    support direct application of Lemma~R16.  Instead I supply the
    following argument---a DT analogue (notably missing from RR) of
    Lemma~R8.
    
    Let the ITT nodes be labelled $A_1,\ldots,A_k$ following the
    topological order, with associated intervention indicator nodes
    $F_1,\ldots,F_k$, respectively.  For $r\leq s$, $F_{r:s}$ will
    denote the sequence $(F_r,\ldots,F_s)$, \etc\@ Define
    $Z_r = (\pre(A_r), A_r)$.
    \begin{lem}
      \label{lem:42}
      For $r=1,\ldots,k$,
      \begin{equation}
        \label{eq:4}
        H_r: \ind {Z_r} {F_{r:k}} {\check F_{1:r-1}}.
      \end{equation}
    \end{lem}
    
    \begin{proof}
      We proceed by backwards induction.
      
      From Definition~R18, we have
      \begin{equation}
        \label{eq:h1k}
        \ind {Z_k} {\check F_k} {\check F_{1:k-1}}.
      \end{equation}
      Then by \lemref{rr16} we have
      \begin{equation}
        \label{eq:hk2}
        \ind {Z_k} {F_k} {\check F_{1:k-1}},
      \end{equation}
      So $H_k$ holds.
      
      Now suppose $H_{r+1}$ holds.  Then
      \begin{equation}
        \label{eq:marg}
        \ind {Z_r} {F_{r+1:k}} {\check F_{1:r}}.
      \end{equation}
      
      Also from Definition~R18, we have
      \begin{equation}
        \label{eq:hr1}
        \ind  {Z_r} {\check F_{r}} {(\check F_{1:r-1}, \check F_{r+1:k})}.
      \end{equation}
      
      
      Now fix a value $\check f\neq\idle$ of $F_{r+1:k}$ (and thus of
      $\check F_{r+1:k}$).  By \eqref{marg}, for any possible
      values $\check f_{1:r}$ of $\check F_{1:r}$ and
      $f_{r+1:k}$ of $F_{r+1:k}$,\footnote{\label{fn:1}Here, and in other similar arguments, we rely on
        \remref{1}.}
      \begin{equation}
        \label{eq:5}
        \Pr(Z_r=z_r \mid\check F_{1:r}=\check f_{1:r},  F_{r+1:k}=f_{r+1:k}) = \Pr(Z_r=z_r \mid\check F_{1:r}=\check f_{1:r},
        \check F_{r+1:k}
        =  \check f).
      \end{equation}
      Since, by \eqref{hr1}, the right-hand side of \eqref{5} is a function
      only of $\check f_{1:r-1}$, the same holds for the left-hand side.
      That is,
      \begin{equation}
        \label{eq:thatisx}
        \ind  {Z_r} {(\check F_{r},  F_{r+1:k})} {\check F_{1:r-1}}.
      \end{equation}
      
      Then $H_r$ follows from \corref{aa} and \corref{bb}, and the induction
      is established.
    \end{proof}
    
    The following is immediate by marginalisation of \eqref{4} (where by
    $F_S$ when $S\not\subseteq A$ we understand  $F_{A\cap S}$):
    
    \begin{cor}
      \label{cor:rr48} \lemref{42} continues to hold if we replace $Z_r$
      by $(W_i,W_{\pre(i)})$, where $W_i$ lies between $A_{r-1}$ and $A_r$
      in the topological order.  That is to say,
      \begin{equation}
        \label{eq:11}
        \ind {(W_i,W_{\pre(i)})} {F_{A\setminus\pre(i)}} {\check F_{\pre(i)}}.
      \end{equation}
      
    \end{cor}
    
    Conditioning on $W_{\pre(i)}$ in \eqref{11}, we deduce the equality of
    (R47) and (R48).
    
  \item[(R49)]
    
    Contrary to RR's assertions,
    \begin{equation}
      \label{eq:13}
      \ind {W_i} {F_{(A\cap\pre(i))\setminus\pa(i)}} {F_{A\cap\pa(i)},F_{A\cap\pre(i)}\neq\emptyset}
    \end{equation}
    does not follow from the local Markov property, and even if it were
    valid would not allow application of Lemma~R17.\\
    
    Instead I prove the following result, which implies the equivalence
    of (R47) and (R49).
    
    \begin{lem}
      \label{lem:rr49}
      \begin{equation}
        \label{eq:rr49}
        \ind{W_i} {F_{A\setminus\parents{i}}} {( W_{\pre(i)},\check F_{\parents{i}})}.
      \end{equation}
    \end{lem}
    
    \begin{proof}
      
      It does follow from the local Markov property that
      \begin{equation}
        \label{eq:15}
        \ind {W_i} {\check F_{\pre(i)\setminus\pa(i)}}
        {W_{\pre(i)}, \check F_{\pa(i)}, \check F_{A\setminus\pre(i)}}, 
      \end{equation}
      Also, by \eqref{11},
      \begin{equation}
        \label{eq:12}
        \ind {W_i} {F_{A\setminus\pre(i)}} {W_{\pre(i)}, \check
          F_{\pa(i)}, \check F_{\pre(i)\setminus\pa(i)}}.
      \end{equation}
      
      Again, fix a value $\check f \neq\emptyset$ of
      $F_{A\setminus\pre(i)}$ (and so of
      $\check F_{A\setminus\pre(i)}$).  By \eqref{12},
      \begin{eqnarray}
        \nonumber
        \lefteqn{\Pr(W_i = w \mid W_{\pre(i)}, \check
        F_{\pa(i)}, \check F_{\pre(i)\setminus\pa(i)},
        F_{A\setminus\pre(i)})}\\
        &=&
            \label{eq:14}
            \Pr(W_i = w \mid W_{\pre(i)}, \check
            F_{\pa(i)}, \check F_{\pre(i)\setminus\pa(i)}, \check F_{A\setminus\pre(i)}=\check f).
      \end{eqnarray}
      By \eqref{15} the right-hand side of \eqref{14} depends only on
      $(W_{\pre(i)}, \check F_{\pa(i)})$.  Then the same holds for the
      left-hand side, proving \eqref{rr49}.
    \end{proof}
    
    Similar arguments deliver (R50) and (R51).
  \end{description}

\section{Discussion}
\label{sec:disc}

RR have very nicely demonstrated the close connexions between their
SWIG approach and my own DT approach, as described in D21.  Minor
notational issues aside, there are two main differences:

\begin{enumerate}
\item While both approaches introduce ``intention-to-treat'' variables
  as a way of justifying ignorability assumptions, once this has been
  done DT can dispense with them, relying on regime indicator
  variables to express and manipulate those assumptions.  SWIGs, on
  the other hand, retain these additional ITT variables explicitly.  I
  contend that the DT approach with regime indicators makes for
  cleaner representation and analysis.
\item A more substantial difference is that SWIGs explicitly represent
  potential outcomes, whereas---as demonstrated by RR, as well as in
  D21--- in a DT analysis they are not needed.
\end{enumerate}

In many articles over many years I have argued convincingly (at least
to my own satisfaction) that the potential outcome approach to
statistical causality is misguided and misleading.  But I can not deny
that---for some unfathomable reason---it is still regarded as
fundamental by most researchers in the field.  It is thus an important
advantage of SWIG representations that they engage more directly with
the large audience of potential outcome enthusiasts.  But I hope that
RR's own clear demonstration that there is a cleaner,
decision-theoretic, way of framing the same problems, in which
potential responses simply have no place, will help to curb that
misplaced enthusiasm.

\section*{Acknowledgments}
\label{sec:ack}
I am grateful to Thomas Richardson and Jamie Robins for continuing
valuable discussions, and to Andrew Yiu for helpful comments.

\bibliographystyle{unsrtnat} 
\bibliography{strings,causal,allclean,ci}

\begin{thebibliography}{15}
\providecommand{\natexlab}[1]{#1}
\providecommand{\url}[1]{\texttt{#1}}
\expandafter\ifx\csname urlstyle\endcsname\relax
  \providecommand{\doi}[1]{doi: #1}\else
  \providecommand{\doi}{doi: \begingroup \urlstyle{rm}\Url}\fi

\bibitem[Richardson and Robins(2023)]{RandR:discussion}
Thomas~S. Richardson and James~M. Robins.
\newblock Potential outcomes and decision theoretic foundations for statistical
  causality.
\newblock \emph{Journal of Causal Inference}, 2023.
\newblock {\bf Details to be
  inserted}\\\href{https://arxiv.org/abs/2302.03899}{\tt arXiv:2302.03899}.

\bibitem[Dawid(2021)]{apd:found}
A.~Philip Dawid.
\newblock Decision-theoretic foundations for statistical causality.
\newblock \emph{Journal of Causal Inference}, 9:\penalty0 39--77, 2021.
\newblock \\{\small\href{http://dx.doi.org/10.1515/jci-2020-0008}{\tt
  DOI:10.1515/jci-2020-0008}}.

\bibitem[Pearl and Mackenzie(2018)]{pearl:why}
Judea Pearl and Dana Mackenzie.
\newblock \emph{The Book of Why}.
\newblock Basic Books, New York, 2018.

\bibitem[Dawid(2007)]{apd:kent}
A.~Philip Dawid.
\newblock Counterfactuals, hypotheticals and potential responses: A
  philosophical examination of statistical causality.
\newblock In Federica Russo and Jon Williamson, editors, \emph{Causality and
  Probability in the Sciences}, volume~5 of \emph{{Texts in Philosophy}}, pages
  503--32. College Publications, London, 2007.

\bibitem[Dawid and Musio(2022)]{apd/mm:eoccoe}
A.~Philip Dawid and Monica Musio.
\newblock Effects of causes and causes of effects.
\newblock \emph{Annual Review of Statistics and its Application}, 9:\penalty0
  261--287, 2022.
\newblock \\
  \href{https://doi.org/10.1146/annurev-statistics-070121-061120}{\tt
  DOI:10.1146/annurev-statistics-070121-06112}.

\bibitem[Holland(1986)]{pwh:jasa}
Paul~W. Holland.
\newblock Statistics and causal inference (with {D}iscussion).
\newblock \emph{Journal of the American Statistical Association}, 81:\penalty0
  945--970, 1986.

\bibitem[Dawid(2000)]{apd:cinfer}
A.~Philip Dawid.
\newblock Causal inference without counterfactuals (with {D}iscussion).
\newblock \emph{Journal of the American Statistical Association}, 95:\penalty0
  407--448, 2000.

\bibitem[Robins(1987)]{jr:cma}
James~M. Robins.
\newblock Addendum to ``{A} new approach to causal inference in mortality
  studies with sustained exposure periods---{A}pplication to control of the
  healthy worker survivor effect''.
\newblock \emph{Computers \& Mathematics with Applications}, 14:\penalty0
  923--945, 1987.

\bibitem[Dawid and Didelez(2010)]{dd:ss}
A.~Philip Dawid and Vanessa Didelez.
\newblock Identifying the consequences of dynamic treatment strategies: A
  decision-theoretic overview.
\newblock \emph{Statistical Surveys}, 4:\penalty0 184--231, 2010.

\bibitem[Dawid(1979)]{apd:misl}
A.~Philip Dawid.
\newblock Some misleading arguments involving conditional independence.
\newblock \emph{Journal of the Royal Statistical Society, Series~B},
  41:\penalty0 249--52, 1979.

\bibitem[Dawid(1980)]{apd:ciso}
A.~Philip Dawid.
\newblock Conditional independence for statistical operations.
\newblock \emph{Annals of Statistics}, 8:\penalty0 598--617, 1980.

\bibitem[Geneletti and Dawid(2011)]{sgg/apd:ett}
Sara~G. Geneletti and A.~Philip Dawid.
\newblock Defining and identifying the effect of treatment on the treated.
\newblock In Phyllis~M. Illari, Federica Russo, and Jon Williamson, editors,
  \emph{Causality in the Sciences}, pages 728--749. Oxford University Press,
  2011.

\bibitem[B\"uhlmann(2020)]{bhlmann2018invariance}
Peter B\"uhlmann.
\newblock Invariance, causality and robustness (with {D}iscussion).
\newblock \emph{Statistical Science}, 35:\penalty0 404--436, 2020.

\bibitem[Pearl and Bareinboim(2014)]{pearl/bareinboim:sts}
Judea Pearl and Elias Bareinboim.
\newblock External validity: From do-calculus to transportability across
  populations.
\newblock \emph{Statistical Science}, 29:\penalty0 579--595, 2014.

\bibitem[Dawid(2001)]{apd:varind}
A.~Philip Dawid.
\newblock Some variations on variation independence.
\newblock In Tommi Jaakkola and Thomas~S. Richardson, editors, \emph{Artificial
  Intelligence and Statistics 2001}, pages 187--191, San Francisco, California,
  2001. Morgan Kaufmann Publishers.

\end{thebibliography}
  

\end{document}